\documentstyle[11pt,amssymb,cite]{article}

\def\ben{\begin{equation}}
\def\een{\end{equation}}

  \let\n=\nu \let\x=\xi

\let\C=\Chi

\def\nn{\nonumber} \def\bd{\begin{document}} \def\ed{\end{document}}
\def\ds{\documentstyle} \let\fr=\frac \let\bl=\bigl \let\br=\bigr
\let\Br=\Bigr \let\Bl=\Bigl
\let\bm=\bibitem
\let\na=\nabla
\let\pa=\partial \let\ov=\overline
\newcommand{\be}{\begin{equation}}
\newcommand{\ee}{\end{equation}}
\def\ba{\begin{array}}
\def\ea{\end{array}}
\def\ft#1#2{{\textstyle{{\scriptstyle #1}\over {\scriptstyle #2}}}}
\def\fft#1#2{{#1 \over #2}}
\def\del{\partial}
\def\vp{\varphi}
\def\sst#1{{\scriptscriptstyle #1}}
\def\oneone{\rlap 1\mkern4mu{\rm l}}
\def\td{\tilde}
\def\wtd{\widetilde}
\def\ie{\rm i.e.\ }
\def\dalemb#1#2{{\vbox{\hrule height .#2pt
        \hbox{\vrule width.#2pt height#1pt \kern#1pt
                \vrule width.#2pt}
        \hrule height.#2pt}}}
\def\square{\mathord{\dalemb{6.8}{7}\hbox{\hskip1pt}}}
\newcommand{\ho}[1]{$\, ^{#1}$}
\newcommand{\hoch}[1]{$\, ^{#1}$}
\newcommand{\bea}{\begin{eqnarray}}
\newcommand{\eea}{\end{eqnarray}}
\newcommand{\ra}{\rightarrow}
\newcommand{\lra}{\longrightarrow}
\newcommand{\Lra}{\Leftrightarrow}
\newcommand{\ap}{\alpha^\prime}
\newcommand{\bp}{\tilde \beta^\prime}
\newcommand{\tr}{{\rm tr} }
\newcommand{\Tr}{{\rm Tr} }
\def\0{{\sst{(0)}}}
\def\1{{\sst{(1)}}}
\def\2{{\sst{(2)}}}
\def\3{{\sst{(3)}}}
\def\4{{\sst{(4)}}}
\def\5{{\sst{(5)}}}
\def\6{{\sst{(6)}}}
\def\7{{\sst{(7)}}}
\def\8{{\sst{(8)}}}
\def\n{{\sst{(n)}}}
\def\cA{{{\cal A}}}
\def\cB{{{\cal B}}}
\def\cF{{{\cal F}}}
\def\cH{{{\cal H}}}
\def\tV{\widetilde V}
\def\tW{\widetilde W}
\def\tH{\widetilde H}
\def\tE{\widetilde E}
\def\tF{\widetilde F}
\def\tA{\widetilde A}
\def\im{{i}}
\def\tY{{{\wtd Y}}}
\def\ep{{\epsilon}}
\def\vep{{\varepsilon}}
\def\R{\rlap{\rm I}\mkern3mu{\rm R}}
\def\bD{{{\bar D}}}

\def\R{\rlap{\rm I}\mkern3mu{\rm R}}
\def\bD{{{\bar D}}}
\def\R{{{\Bbb R}}}
\def\C{{{\Bbb C}}}
\def\H{{{\Bbb H}}}
\def\CP{{{\Bbb C}{\Bbb P}}}
\def\RP{{{\Bbb R}{\Bbb P}}}
\def\Z{{{\Bbb Z}}}
\def\bA{{{\Bbb A}}}
\def\bB{{{\Bbb B}}}
\def\bC{{{\Bbb C}}}
\def\bD{{{\Bbb D}}}
\def\bE{{{\Bbb E}}}
\def\bZ{{{\Bbb Z}}}
\def\Re{{{\frak{Re}}}}
\def\Im{{{\frak{Im}}}}
\def\cosec{{\,\hbox{cosec}\,}}
\def\Gm{{\Gamma_{\!\! -}}}
\def\Gp{{\Gamma_{\!\! +}}}
\def\stan{{standard }}
\def\nonstan{{supernumerary }}
\def\FF2{{ {}_{\sst 2}F_{\sst 1} }}

\newcommand{\tamphys}{\it Center for Theoretical Physics,
Texas A\&M University, College Station, TX 77843}

\newcommand{\upenn}{\it Department of Physics and Astronomy,\\ University
of Pennsylvania, Philadelphia, PA 19104}

\newcommand{\brussels}{\it Physique Th\'eorique et Math\'ematique,
Universit\'e Libre de Bruxelles,\\ Campus Plaine C.P. 231, B-1050
Bruxelles, Belgium}

\newcommand{\auth}{Z.-W. Chong\,, 
H. L\"u and C.N. Pope}

\begin{document}
\begin{flushright}

MIFP-04-03\\
{\bf hep-th/0402202}\\
February\  2004
\end{flushright}

\vspace{10pt}

\begin{center}

{\large {\bf Rotating Strings in Massive Type IIA Supergravity}}

\vspace{20pt}
\auth

\vspace{20pt}

{\it George P. and Cynthia W. Mitchell
Institute for Fundamental Physics,\\ Texas A\& M University,
College Station, TX 77843-4242, USA}

%\vspace{10pt} {\hoch{\dagger}\brussels}

\vspace{40pt}

\underline{ABSTRACT}
\end{center}

    Massive type IIA supergravity admits a warped AdS$_6\times S^4$
vacuum solution, which is expected to be dual to an ${\cal N}=2$, $D=5$
super-conformal Yang-Mills theory.  We study solutions for strings
rotating or spinning in this background.  The warp factor plays no
essential role when the string spins in the AdS$_6$, implying a
commonality in the leading Regge trajectories between the $D=4$ and
$D=5$ super-conformal field theories.  The warp factor does, however,
become important when the string rotates in the $S^4$, in particular
for long strings, which have the the relation $E-\ft32 J=c_1 +
c_2/J^5+\cdots$, where the angular momentum $J$ is large.  This
relation is qualitatively different from that for long strings in the
AdS$_5\times S^5$ background.  We also study Penrose limits of the
AdS$_6\times S^4$ solution, one of which gives rise to a free massive
string theory with time-dependent masses.

{\vfill\leftline{}\vfill \vskip 10pt \footnoterule 
{\footnotesize
Research supported in part by DOE grant
DE-FG03-95ER40917.}

}

\pagebreak
\setcounter{page}{1}

%\tableofcontents
%\addtocontents{toc}{\protect\setcounter{tocdepth}{2}}
\newpage

\section{Introduction}

      Spinning extended object solutions, such as spherical membranes,
were first constructed in supergravity some time ago 
\cite{membrane,membrane2,membrane3}.
The spin is necessary in order to prevent the membrane from collapsing.
Inspired by the AdS/CFT correspondence, spinning string solutions were
obtained in the AdS$_5\times S^5$ background.  These solutions correspond
to string states on the leading Regge trajectory, with large angular
momentum \cite{gkp,ft}.  Subsequently, there has been a considerable
interest in the subject [6-18]. (See a recent review \cite{treview}.)

     The majority of the focus has been on the AdS$_5\times S^5$ and
AdS$_5\times T^{1,1}$ backgrounds of the type IIB theory.  One nice
feature of these solutions is that they are supported only by a
Ramond-Ramond 5-form field strength, and so the bosonic sector of the
sigma model action for the background is rather simple, with no
Wess-Zumino term.

     In this paper, we consider the warped AdS$_6\times S^4$
background of the massive type IIA supergravity \cite{romans10}.  It
is expected to be dual \cite{ferrarai,ads6} to an ${\cal N}=2$, $D=5$
superconformal Yang-Mills theory \cite{sei,int}.  This solution is
also supported just by Ramond-Ramond field strengths, but now in
addition with a non-constant dilaton.  The AdS$_6\times S^4$ metric is
warped, rather than a direct product, with a warp factor that becomes
singular on the equator of $S^4$, and so the geometry really
corresponds to a hemisphere instead of the full $S^4$.  The string
coupling diverges at the equator; the metric is singular in the
Einstein frame, but regular in the string frame.\footnote{For brevity,
we shall generally refer to the solution as AdS$_6\times S^4$, with
the understanding that the 4-sphere is cut at the
equator.}\label{foot1} The warp factor implies that the equator of the
$S^4$ corresponds to an AdS$_5\times S^3$ boundary \cite{ads6}.

   We shall study spinning and rotating strings in this AdS$_6\times S^4$
background.  We consider two types of such solution.  First, we
study strings rotating in the $S^4$, for which the warp factor can be 
important.  As one might expect, for a short string the warp factor
plays little role, but its effect becomes pronounced for long strings
with large R-charge $J$.  The energy $E$ and angular momentum $J$ are
now related by
%%%%
\be
E-\ft32 J \sim c_1 + \fft{c_2}{J^5} + \cdots
\ee
%%%%%
where $c_1$ and $c_2$ are certain constants.  This behaviour is
qualitatively different from the analogous relation for strings 
spinning in an AdS$_5\times S^5$ background.

      In the AdS$_5\times S^5$ background, one can consider a string
rotating in an $S^3$ inside the $S^5$ \cite{Frolov:2003xy}.  The $S^3$
lies within a great circle (or, more precisely, a ``great 4-sphere'')
in the $S^5$.  Were it not for the warp factor, an analogous solution
would also be possible in AdS$_6\times S^4$, with the $S^3$ inside the
$S^4$.  However, the only ``great 3-sphere'' in the northern
hemisphere of $S^4$ is the equator itself, which is precisely where
the warp factor diverges, and for this reason the analogous solution
does not arise.

       We also study situations where the string spins purely in the
AdS$_6$ spacetime.  In this case, the warp factor plays no essential
role, since the gravitional repulsion from the boundary at the equator
implies that the string can only be located at the north pole.  We
obtain results that are identical to those for a string spinning in
AdS$_5$ in the type IIB theory, implying a commonality of the leading
Regge trajectories for the $D=5$ and $D=4$ theories.

    We then study two different Penrose limits of the AdS$_6\times
S^4$ solution.  In the first of these, we arrive at a pp-wave which
gives rise to a massive string theory with time-dependent masses.  The
singularity of the warp factor on the great 3-sphere in $S^4$ is
reached in a finite string worldsheet time $x^+=x^+_0$.  The dilaton
also depends on the worldsheet time, and in fact the string coupling
becomes infinite at this limiting time $x^+_0$.  This may signal a
breakdown of the validity of the background.  We also consider a
second Penrose limit, which gives rise to an interacting string
theory.  Had there not been a warp factor in the AdS$_6\times S^4$
solution, these two Penrose limits would have been equivalent.

\section{General equations}

The massive type IIA supergravity theory supports a warped AdS$_6\times S^4$
background, which arises as the near-horizon geometry \cite{ads6} of
a semi-localised D4-D8 system \cite{d4d8}.   The solution is given by
%%%%%
\bea
ds^2 &=& \ft12 W(\xi)^2\,\Bigl[
9(-\cosh^2\rho\, dt^2 + d\rho^2 + \sinh^2\rho\, d\Omega_4^2)
+4(d\xi^2 + \sin^2\xi\, d\Omega_3^2)\Bigr]\nn\\
F_\4 &=& \ft{20\sqrt2}{3}\, (\cos\xi)^{\ft13}\,\sin^3\xi\, 
d\xi\wedge\Omega_\3\,,\qquad
e^{\phi} = (\cos\xi)^{-\ft56}\,,\label{ads6s4}
\eea
where $d\Omega_4^2$ and $d\Omega_3^2$ are the metrics for the unit
$S^4$ and $S^3$, which we choose to parameterise as follows:
%%%%
\bea
d\Omega_4^2 &=& d\theta_1^2 + \cos\theta_1^2 \, (
d\theta_2^2 + \cos^2\theta_2\, d\phi_1^2 +
\sin^2\theta_2\, d\phi_2^2)\,,\nn\\
d\Omega_3^2 &=& d\theta^2 + \cos^2\theta\, d\psi_1^2 +
\sin^2\theta\, d\psi_2^2\,,
\eea
%%%%
and $\Omega_\3$ is the volume form of $d\Omega_3^2$.  The solution
preserves half of the maximal supersymmetry, and remarkably, it
corresponds to the vacuum of a consistent Kaluza-Klein reduction on
the warped $S^4$ that gives rise \cite{adsred} to the $D=6$, ${\cal
N}=2$ gauged supergravity constructed by Romans \cite{romans6}.

      The warp factor $W(\xi)$ in the metric (\ref{ads6s4}) is given
by $W=(\cos\xi)^{-\ft{1}{6}}$ in the string frame, and
$W=(\cos\xi)^\ft1{24}$ in the Einstein frame. 
The coordinate $\xi$ runs from 0 to $\ft12\pi$, at which the metric in
the Einstein frame becomes singular, while in the string frame the
metric remains regular there.  To be precise, the metric describes a warped
product of the AdS$_6$ with the upper hemisphere of the $S^4$.  Since
we will use this solution as background geometry in the string
sigma-model action, we should work with the string-frame metric.  The
equator $\xi= \ft12\pi$ can be viewed as an AdS$_6\times S^3$
boundary.

     It is straightforward to write down the bosonic sector of the
string $\sigma$-model action, since the ``cosmological term'' and the
4-form $F_\4$ are R-R fields, which enter the action only through
fermion bilinears, as does the dilaton. Choosing the conformal gauge,
we can make the following consistent ansatz that describes strings
spinning in the $AdS_6\times S^4$ background:
%%%
\bea
&&t=\kappa\, \tau\,,\qquad \psi_1=\omega_1\, \tau\,,\qquad
\psi_2=\omega_2\, \tau\,,\qquad \phi_1=\omega_3\,\tau\,,\qquad
\phi_2=\omega_4\,\tau\nn\\
&&\rho=\rho(\sigma)\,,\qquad \xi=\xi(\sigma)\, \qquad
\theta=\theta(\sigma)\,,\qquad \theta_1=\theta_1(\sigma)\,,\qquad
\theta_2=\theta_2(\sigma)\,,
\eea
%%%
where a prime denotes a derivative with respect to the world-sheet
coordinate $\sigma$.  The action becomes
%%%
\bea
I = \fft{1}{4\pi\,\alpha'}\, \int d\sigma\, {\cal L}
\eea
%%%%
where
%%%%
\bea
{\cal L}&=& 9W^2\, \Bigl(-\kappa^2\, \cosh^2\rho - \rho'^2 +
\sinh^2\rho\, (-\theta_1'^2 -\cos^2\theta_1\, \theta_2'^2
\nn\\
&&\qquad\qquad\qquad + 
\omega_3^2\, \cos^2\theta_1\,\cos^2\theta_2 +
\omega_4^2\, \cos^2\theta_1\,\sin^2\theta_2) \Bigr)\nn\\
&&+ 4W^2\,  \Bigl(-\xi'^2 + \sin^2\xi\, (-\theta'^2 +
\omega_1^2\, \cos^2\theta + \omega_2^2\,\sin^2\theta)\Bigr)\,,
\eea
%%%%%
together with the conformal gauge constraint
%%%%
\bea
&&9W^2\, \Bigl(-\kappa^2\, \cosh^2\rho + \rho'^2 +
\sinh^2\rho\, (\theta_1'^2 +\cos^2\theta_1\, \theta_2'^2
\nn\\
&&\qquad\qquad\qquad +
\omega_3^2\, \cos^2\theta_1\,\cos^2\theta_2 +
\omega_4^2\, \cos^2\theta_1\,\sin^2\theta_2) \Bigr)\nn\\
&&+ 4W^2\,  \Bigl(\xi'^2 + \sin^2\xi\, (\theta'^2 +
\omega_1^2\, \cos^2\theta + \omega_2^2\,\sin^2\theta)\Bigr)= 0\,,
\eea
%%%%
where $W\equiv (\cos\xi)^{-1/6}$.
The equations of motion for this system are given by
%%%%
\bea
&&(W^2\, \rho')' + \ft12 W^2\, \sinh2\rho\, \Bigl(
-\kappa^2 - \theta_1'^2 + \cos^2\theta_1\, (
-\theta_2'^2 + \omega_3^2\, \cos^2\theta_2 +
\omega_4^2\, \sin^2\theta_2\Bigr) =0\,,\nn\\
&&(W^2\, \sinh^2\rho\, \theta_1')' -
\ft12 W^2\, \sinh^2\rho\, \sin(2\theta_1)\, (-\theta_2'^2 +
\omega_3^2\, \cos^2\theta_2 + \omega_4^2\, \sin^2\theta_2) =0\,,\nn\\
%%%
&&(W^2\, \sinh^2\rho\, \cos^2\theta_1\, \theta_2')' +
\ft12 W^2\, \sinh^2\rho\,\cos^2\theta_1\, \sin2\theta_2
(\omega_4^2 - \omega_3^2) =0\,,\nn\\
%%%
&& (W^2\, \xi')' + \ft1{24} \tan\xi \, 
{\cal L} +
\ft12W^2\, \sin2\xi\, (-\theta'^2 + \omega_1^2\, \cos\theta^2 +
\omega_2^2\, \sin^2\theta)=0\,,\nn\\
%%%
&&(W^2\,\sin^2\xi\,\theta')' + \ft12 W^2\, 
\sin^2\xi\, \sin2\theta\, (\omega_2^2 - \omega_1^2)=0
\,.
\eea
%%%%
It is unlikely that one can obtain the most general solutions to these
equations explicitly, especially since the warp factor $W$ introduces
fractional powers of $\cos\xi$.  We shall consider restricted cases
in which analytical solutions can be obtained.

\section{Rotation in the warped $S^4$}

    We first consider rotations in the warped $S^4$.  The angular
momentum in this case can be related to the R-charges of operators in
the corresponding $D=5$ conformal field theory.  There exist a maximum
of two communting $U(1)$ isometries in $S^4$.  Here, we shall consider
just one angular momentum, by setting the parameters associated with
these two charges equal.  To turn off the contribution from the AdS$_6$
directions, we can either set $\rho=0$, or set $\rho$ to be a constant
and take $\kappa=\omega_3=\omega_4$; both yield the same result.  In the
$S^4$ direction, we set $\omega_2=\omega_1\equiv\omega$, which implies
%%%%
\be
\theta' = \fft{c}{W^2\, \sin^2\xi}\,.
\ee
%%%
Substituting this into the the constraint equation, we have
%%%
\be
\xi'^2 = \ft94 \kappa^2 - \omega^2\, \sin^2\xi - \fft{c^2}{W^4\,
\sin^2\xi}\,.
\ee
%%%%%
When $c\ne 0$, $\xi$ has two turning points, $0< \xi_1 <\x_2$, but
when $c=0$ there is
only one turning point $\xi_2$, and the coordinate
$\xi$ oscillates between $-\xi_2$ and $\xi_2$.  
\footnote{The hemisphere of $S^4$ is normally
specified by $0\le\xi\le\ft12\pi$.  When the oscillating solution passes into 
the region with $\xi<0$, this should really be re-interpreted via 
a coordinate transformation under which $\xi\longrightarrow -\xi$ 
together with an antipodal mapping on the $S^3$ foliating surfaces.  
This is analogous to re-interpreting motion into the region $\theta<0$  
along a polar great circle on $S^2$ with standard spherical polar coordinates
$(\theta,\phi)$ via the coordinate transformation
$(\theta,\phi)\longrightarrow (-\theta,\phi+\pi)$.}\label{foot2} 
There does not seem 
to exist an analytical
solution with $c\ne 0$, and so we shall specialise to the case $c=0$.
The parameters $\kappa$ and $\omega$ then satisfy the
constraint
%%%%
\be
4\int_0^{\xi_0} \fft{d\xi}{\sqrt{\ft94\kappa^2 -\omega^2\, \sin^2\xi}}
=\int d\sigma=2\pi\,.\label{s4con}
\ee 
%%%%%
Defining $\eta=2\omega/(3\kappa)$, then for $\eta\ge 1$ there exists
a turning point $\xi_0$ given by $\sin\xi_0=1/\eta$. 
Note that $\xi$ is a compact coordinate,
running from $0$ to $\ft12\pi$.  Clearly, if $\eta >>0$ then $\xi_0$ is 
close to 0, and we have a short string.  On the other hand, if
$\eta \sim 1^+$  we have $\xi_0$ close to $\pi/2$, and the string is ``long''.
In terms of $\eta$, the constraint (\ref{s4con}) reads
%%%
\be
\eta^{-\ft12}\, \FF2[\ft12, \ft12, 1; \eta^{-1}] = \ft32\kappa
\ee
%%%

    The energy and the angular momentum are given by
%%%%
\bea
E &=&\int\fft{d\sigma}{2\pi}\, P_t = \fft{12}{\pi}
\int_0^{\xi_0} (\cos\xi)^{-1/3}\, (1-\eta\, \sin^2\xi)^{-1/2}\nn\\
&=& \fft{6}{\sqrt\eta}\, \FF2[\ft23, \ft12, 1; \eta^{-1}]\,,\nn\\
J &=& \int\fft{d\sigma}{2\pi}\, P_\phi =
\fft{2\sqrt\eta}{\pi} \int_0^{\xi_0}
d\xi\, (\cos\xi)^{-1/3}\, \sin^2\xi\, (1-\eta\, \sin^2\xi)^{-1/2}\nn\\ 
&=&\fft{2}{\eta}\, \FF2[\ft23, \ft32, 2; \eta^{-1}]
\eea
%%%%%
We can easily determine the form of the relation between energy and
spin in each of the short and the ``long'' string limits.

   When $\eta>>1$, implying a short string, both $E$ and $J$ approach
zero. They obey the relation
%%%
\be
E=3\sqrt{2J}\, (1 + \ft1{24} J + \ft1{1152} J^2 -
\ft7{82944} J^3 + \cdots)\,.\label{ej1}
\ee
%%%%
If, on the other hand, $\eta^{-1}=1-\epsilon$ with $\epsilon$
approaching zero from above, the string is long.  In this case, both $E$
and $J$ approach infinity with power-law dependences on $\epsilon^{-1}$, 
and $E$ and $J$ are related by
%%%%
\be
E-\ft32 J =  \fft{6\Gamma(\ft56)}{\sqrt\pi\, \Gamma(\ft43)} -
\fft{3072\Gamma(\ft16)^6}{5\pi^3\, \Gamma(\ft23)^6\, J^5}
+\fft{2^{\ft13}\, 82944\,   \Gamma(\ft16)^6}{\pi^5\, 
\Gamma(\ft23)^3\, J^6} + \cdots\,.
\ee
%%%
An alternative expression for the above relation is
%%%
\bea
E-\fft3{2\sin\xi_0}\, J &=& E - \ft32\sqrt\eta\, J =
\fft{3}{\sqrt{\eta}}\, \FF2[\ft23, \ft12, 2; \eta^{-1}]\nn\\
&=& \fft{6\Gamma(\ft56)}{\sqrt\pi\, \Gamma(\ft43)}\, 
(1 + \ft32 \epsilon + \ft{57}{56}\, \epsilon^2 + \cdots)\nn\\
&& +
\fft{3\Gamma(-\ft56)}{\sqrt\pi\, \Gamma(\ft23)}\, \epsilon^{\ft56}
\, (1 + \ft{13}{22}\epsilon + \ft{677}{1496} \epsilon^2 +
\cdots)\,.
\eea
%%%%
\bigskip

\noindent{\underline{Comparison with AdS$_5\times S^5$}:}
\bigskip

  Having obtained the results for strings rotating in the $S^4$, it is
instructive to compare them with the previously-obtained results for
strings rotating in the $S^5$ in an AdS$_5\times S^5$ background,
where there is no warp factor \cite{gkp,ft}.  The metric of the
$S^5$ can be parameterised as
%%%%
\be
d\Omega_5^2 = d\xi^2 + \sin^2\xi\, \Bigl(d\theta_1^2 + 
\cos\theta_1^2\, (d\theta^2 + \cos^2\theta\, d\phi_1^2 +
\sin^2\theta\, d\phi_2^2)\Big)\,.
\ee
%%%
To make a direct comparison with our warped $S^4$ result, we consider
the string spinning in an $S^4$ section ($\theta_1=0$) of the $S^5$.
The solution is given by $t=\kappa\, \tau$, $\phi_1=\omega\, \tau=
\phi_2$, $\theta={\rm const.}$, where $\xi=\xi(\sigma)$ satisfies the
contraint equation
%%%%
\be
\xi'^2 = \kappa^2 - \omega^2\, \sin^2\xi\,.
\ee
%%%%
This implies that $\kappa$ and $\omega$ are related by
%%%%%
\be
\kappa = \fft1{\sqrt\eta}\, \FF2[\ft12,\ft12,1;\eta^{-1}] =
\fft{2}{\pi\, \sqrt{\eta}}\, K(\eta^{-1})
\,,
\ee
%%%
where $K$ is the complete elliptic integral of the first kind.
The energy and the angular momentum can be easily obtained, given by
%%%%
\be
E= \kappa= \fft1{\sqrt\eta} \,  \FF2[\ft12,\ft12,1;\eta^{-1}]=
\fft{2}{\pi\, \sqrt{\eta}}\, K(\eta^{-1})\,,\qquad
J= \fft1{2\eta}\, \FF2[\ft12, \ft32, 2; \eta^{-1}]\,,
\ee
%%%%
For short strings, where $\eta>>1$, the energy and spin both approach
to zero, and they obey the relation
%%%%
\be
E=\sqrt{2J}\, (1 + \ft18J + \ft3{128}\, J^2 + \ft{1}{1024} J^3 +\cdots)\,.
\label{ej2}
\ee
%%%%%
For a ``long'' string where $\eta^{-1}\sim 1-\epsilon$,
$E$ and $J$  both diverge logarithmically as $\log\epsilon$,
while the difference $E-J$ approaches a constant:  
%%%%
\be
E- J= \fft{2}{\pi}  - \fft8{\pi}\, e^{-\pi\, J-2} + \cdots \label{ads5s4}
\ee
%%%%%%

   Thus we see that for short strings, located near the north pole 
where the efffect of any warp
factor is negligible, the energy-angular momentum relations are
qualitatively the same for the warped AdS$_6\times S^4$ and the
AdS$_5\times S^5$ backgrounds.  In fact, (\ref{ej2}) is the same as
(\ref{ej1}), in the first three leading terms, after sending
$J\rightarrow J/3$ and $E\rightarrow E/(3\sqrt3)$.  For long strings, on
the other hand, although at the leading order $E- \wtd\Delta\, J$
approaches a constant for both the AdS$_6\times S^4$ and the AdS$_5\times
S^5$ backgrounds, the next-to-leading order is quite different in the
two cases.  Note that $\wtd\Delta$ is the conformal dimension for the
R-charge operators.

     In the AdS$_5\times S^5$ background, solutions describing
a string rotating in an $S^3$ submanifold of $S^5$ were also obtained
\cite{Frolov:2003xy}.  In our AdS$_6\times S^4$ case, since the $S^4$
can be viewed as a foliation by $S^3$ surfaces, one might expect such a
solution also to exist.  However, owing to a repulsion from the equatorial
boundary implied by the warp factor, any solution with constant lattitude
coordinate $\xi$ necessarily lies at the north pole of the $S^4$, which
renders a rotation only in the $S^3$ impossible.  Thus we see that the
warp factor implies that the energy/R-charge relation obtained in the
$D=4$ Yang-mills theory corresponding to the AdS$_5\times S^5$ background
is significantly different from the one for the $D=5$ Yang-Mills theory
from AdS$_6\times S^4$.

\section{Penrose limit of AdS$_6\times S^4$}

    In the case of the AdS$_5\times S^5$ background, long strings
rotating in $S^5$ correspond to states with large RR-charge, which are 
closely related to pp-waves via the Penrose limit.  The
situation is more complicated if we look for a direct analogue in the
warped AdS$_6\times S^4$ background, owing to the presence of the warp
factor.  We shall first consider the Penrose limit using the method outlined
in \cite{blau}.  

   We begin by expressing the metric (\ref{ads6s4}) (after a 
constant scaling) as
%%%%%
\be
ds^2 = \ft12 \lambda^2\, (\cos\xi)^{-\ft13}\, \Bigl[ 9 (-d\tau^2 +
\sin^2\tau\, d\Omega_{H^5}^2) + 4\, (d\xi^2 + \sin^2\xi\, 
d\Omega_3^2)\Bigr]\,,
\ee
%%%%
where $d\Omega_{H^5}^2$ is a unit hyperbolic 5-plane.  Making the coordinate
transformation $u=\xi + \ft32 \tau$ and $v=\lambda^2\,(\xi - \ft32 \tau)$,
and then sending $\lambda\rightarrow \infty$, we obtain a pp-wave in the
form 
%%%%%
\be
ds^2 = 2( du\, dv + \ft94 \sin^2(\ft13 u)\, d\td y^i d\td y^i + 
          \sin^2(\ft12 u)\, d\td z^i d\td z^i)\,,
\ee
%%%%%
where the metrics $\lambda^2\, d\Omega_{H^5}^2$ and $\lambda^2\, 
d\Omega_3^2$, which become flat in the limit $\lambda\longrightarrow \infty$,
are written as $ d\td y^i d\td y^i$ and $d\td z^m d\td z^m$ respectively.
After the further coordinate transformations (see \cite{blau})
%%%%%
\bea
&& \bar x^+ = \ft12 u\,,\qquad 
\bar x^- = v - \ft14( \ft32 \td y_i^2\, \sin\ft{2}3 u + \td z_m^2\, \sin u)
\,,\nn\\
&& \bar y^i = \ft{3}{\sqrt 2}\, \td y^i\, \sin\ft13 u\,,
\qquad \bar z^m = \sqrt2\, \td z^m\, \sin\ft12 u\,,
\eea
%%%%%
the pp-wave metric takes the form
%%%%%
\be
ds^2 = (\cos \bar x^+)^{-\ft13}\, [4 d\bar x^+\, d\bar x^- 
   - (\ft49 \bar y_i^2 + \bar z_i^2)\, (d\bar x^+)^2  + 
   d\bar y_i^2 + d \bar z_m^2]\,.
\ee
%%%%%
where there are five coordinates $\bar y^i$ and three coordinates $\bar z^m$.
We can now perform the coordinate transformation
%%%%
\bea
&&W^2\, d\bar x^+ = dx^+ \,,\qquad
\bar x^- = x^- + \fft{W'}{4W}\, (y^i\, y^i + z^m\, z^m)\,,\nn\\
&&\bar y_i = \fft{y_i}{W}\,,\qquad
\bar z^m = \fft{z^m}{W}\,,
\eea
%%%
where $W=(\cos\td x^+)^{-\ft16}$, and a prime denotes a derivative
with respect to $x^+$.  The metric becomes
%%%%
\bea
ds^2 = 4 dx^+\, dx^- - \Bigl[(\fft{4}{9W^4} - \fft{W''}{W})\, y^i\, y^i +
(\fft{1}{W^4} - \fft{W''}{W})\, z^m\, z^m\Bigr] +
dy^i\, dy^i + dz^m\, dz^m\,.
\eea
%%%%%%
Thus we see that string theory on this pp-wave background becomes
a massive free string, but with time-dependent masses $m_1$ and $m_2$ in 
the $y^i$ and $z^m$ directions respectively, given by
%%%
\be
m_1^2 = \fft{8 \cos^2 x^+ - 7}{36 \cos^2 x^+} + \ft49 (\cos x^+)^{\ft23}
  \,,\qquad
m_2^2 = \fft{8 \cos^2 x^+ - 7}{36 \cos^2 x^+} + (\cos x^+)^{\ft23}\,.
\ee
%%%
The coordinates $\bar x^+$ and $x^+$ are related by
%%%
\be
x^+ = - \ft32 (\cos\td x^+)^{\ft23}\, \FF2[\ft13, \ft12, \ft43;
\cos^2 \bar x^+]\,.
\ee
%%%%%
The coordinate $\bar x^+$ runs from 0 to $\ft12\pi$, at which point
the string coupling constant $g_s=e^{\phi}=(\cos\bar x^+)^{-5/6}$
becomes infinite.  The coordinate $x^+$ runs
from $-\ft32\sqrt\pi\, \Gamma(\ft43)/\Gamma(\ft56)$ to 0.
Some aspects of massive string theory with time dependent masses
were discussed in \cite{zayas}

      It is also of interest to examine the Penrose limit described in
\cite{bmn}, since this makes a direct analogy with the rotating
strings we obtained earlier.  In this procedure, one magnifies the null
geodesics along a great cirlce of the internal space.  This is rather
problematic in our case, since the great 3-sphere is the 
equator of $S^4$, on which the solution becomes singular.  In order to 
obtain a regular solution, it is necesary to scale the fields
appropriately using
global symmetries of the theory. We begin by making the coordinate
transformations
%%%%
\be
t\rightarrow x^+ + \fft{x^-}{9\lambda^2}\,,\qquad
\psi_1 \rightarrow \ft32 x^+ - \fft{x^-}{6\lambda^2}\,,\qquad
\rho \rightarrow \fft{\rho}{3\lambda}\,,\qquad
\theta\rightarrow \fft{\theta}{2\lambda}\,,\qquad
\xi \rightarrow \ft12\pi - \fft{\xi}{2\lambda}\,.
\ee
%%%%
The solution (\ref{ads6s4}), after a scaling of fields in which the metric is
multiplied by 2 for convenience,  becomes 
%%%%
\be
d\td s^2 = 2^{-\ft23} \lambda^{-\ft53}\, ds^2\,,\qquad
\wtd F_\4 = 2^{-\ft13} \lambda^{-\ft{10}{3}}\, F_\4\,,\qquad
e^{\td\phi} = (2\lambda)^{\ft56}\, e^{\phi}\,,
\label{solscale}
\ee
%%%%%
where
%%%%
\bea
ds^2\!\!\! &=&\!\!\! \xi^{-\ft13}\, \Bigl[ -4 dx^+ dx^- -
[\rho^2 + \ft94(\xi^2 + \theta^2)]\, (dx^+)^2 + d\rho^2 +
\rho^2\, d\Omega_4^2 + d\xi^2 + d\theta^2 + \theta^2\, d\psi_2^2\Bigr]
\,,\nn\\
F_\4\!\!\! &=&\!\!\! 5\xi^{\ft13}\, d\xi\wedge \theta\, d\theta\wedge
d\psi_2\wedge dx^+\,,\qquad
e^{\phi}=\xi^{-\ft56}\,.\label{solresc}
\eea
%%%%

    Normally, when one takes a Penrose limit of AdS$_5\times S^5$, 
AdS$_4\times S^7$ or AdS$_7\times S^4$, one just uses the homogeneous 
global scaling symmetry of the supergravity equations of motion in order to
absorb the singular $\lambda$ scaling factors in the limiting forms 
analogous to (\ref{solscale}).  In the present case, in which the dilaton
is excited and also has a $\lambda$ scaling, it is necessary also to
make use of the dilaton shift symmetry of the type IIA theory, in 
conjunction with the homogeneous scaling symmetry.  These two symmetries
take the form
%%%%%%
\bea
&&
\td g_{\mu\nu} = \kappa^2\, \Lambda^2\, g_{\mu\nu}\,,\qquad
e^{\td\phi}= \Lambda^3\, e^{\phi}\,,\qquad
\wtd F_\3 = \kappa^2\, \Lambda^2\, F_\3\,,\nn\\
&&
\wtd F_\4=\kappa^3\, F_\4\,,\qquad
\wtd F_\2 =\kappa\, \Lambda^{-2}\,F_\2\,,\qquad
\wtd m = \kappa^{-1}\, \Lambda^{-4}\, m\,,
\eea
%%%%
where $\kappa$ and $\Lambda$ are the associated global parameters.  To
absorb the $\lambda$ dependences in (\ref{solscale}) we therefore
choose these parameters to be
%%%
\be
\kappa=2^{-\ft19}\, \lambda^{-\ft{10}{3}}\,,\qquad
\Lambda = (2\lambda)^{\ft5{18}}\,.
\ee
%%%%%
The rescaled pp-wave solution is then given simply by (\ref{solresc}).
The canonical momenta $p^\pm$ in this case can be related
to the rotating string solutions via
%%%%%
\be
p^- = \ft{\im}2\, (\del_t + \ft32 \del_{\psi_1}) = \ft12 (E-\ft32 J)
\,,\qquad
p^+ = \fft{\im}{18\lambda^2}\,  (\del_t - \ft32\del_{\psi_1})=
\fft1{18\lambda^2} (E + \ft32 J)
\ee
%%%%%

\section{String spinning in AdS$_6$}

      Here we consider a string spinning purely in the AdS$_6$
spacetime, in which case $\xi$ is a constant.  In fact the equations
of motion require that $\xi=0$, since there is effectively a
gravitational repulsion because of the warp factor, which forces the
string to be at the north pole.  Since there are only two commuting
$U(1)$ isometries in the AdS$_6$, there can at most be two commuting
angular momenta.  For simplicity, we shall consider only solutions
with one angular momentum parameter, by setting the two angular
momenta equal.  Furthermore, we shall focus on the solution where
$\theta_1=0$, and where we set $\omega_3=\omega_4\equiv\omega$.  This
leads to
%%%%
\be \theta_2' = \fft{c}{\sinh^2\rho}\,,
\ee 
%%%%
where $c$ is a constant.  Substituting this into the conformal constraint,
we obtain 
%%%
\bea \rho'^2 = \kappa^2\, \cosh^2\rho - \omega^2\,
\sinh^2\rho - \fft{c^2}{\sinh^2\rho}\,.
\eea
%%%%
Periodicity in
$\sigma$ implies that we must have
%%%%
\bea n\,\int_{\rho_1}^{\rho_2}
\fft{d\rho}{\sqrt{\kappa^2\, \cosh^2\rho - \omega^2\, \sinh^2\rho -
c^2/\sinh^2\rho}} = \int d\sigma=2\pi\,,
\eea
%%%%%
where $\rho_1$ and $\rho_2$ are the turning points of the oscillatory
solution.  If $c\ne0$, these two turning points both occur at positive
values of $\rho$, and we take $n=2$, since a complete period of the
oscillation runs from $\rho_1$ to $\rho_2$ and then back to $\rho_1$.
If instead $c=0$, the motion runs from the turning point at $\rho=
\rho_2$, passes through zero, and turns again at $\rho=-\rho_2$.  Thus
in this case we can take $\rho_1=0$ (where it is now the mid-point,
rather than a turning point, of the oscillatory motion) and set $n=4$,
since a complete period consists of four segments between $\rho=0$ and
$\rho=\rho_2$.  (See footnote 2.)  By making appropriate
coordinate transformations, the above integral can be expressed as a
hypergeometric function.  To see this, let us define
$t=-\cosh^2\rho/(y_+ - y_-)$, where
%%%%
\be y_\pm =\fft{2\omega^2 - \kappa^2 \pm \sqrt{\kappa^4 + 4c^2\,
(\kappa^2 -\omega^2)}}{2(\omega^2-\kappa^2)}\,.
\ee
%%%%%
For $\omega^2 >
\kappa^2$, we have $y_+ > y_- >1$, ensuring that the corresponding $\rho$
is real.  When $c=0$ we have $y_-=1$, corresponding $\rho=0$, which is
no longer a turning point, and then as discussed above we then take $n=4$
instead of $n=2$.  In terms of the new coordinate $t$ we have
%%%%
\bea 
2\pi &=&
\fft{n}{2\sqrt{y_+\,(\omega^2-\kappa^2)}}\, \int_0^1 \fft{dt}{t^{\ft12}\,
(1-t)^{\ft12}\, (1-\lambda\, t)^{\ft12}} \nn\\ &=& \fft{n\,\pi}{2\,
\sqrt{y_+\, (\omega^2-\kappa^2)}}\, \FF2[\ft12, \ft12, 1; \lambda]\,,
\eea
%%%%%%
where 
%%%
\be 
\lambda = (y_+ - y_-)/y_+\,.  
\ee
%%%%%%
It is straightforward to obtain the energy and angular momentum for
this system, given by
%%%%
\bea E &=& \fft{9n\, \kappa}{2\pi}\, \int
\fft{\cosh^2\rho\, d\rho}{\sqrt{\kappa^2\, \cosh^2\rho - \omega^2\,
\sinh^2\rho - c^2/\sinh^2\rho}}\,,\nn\\
S &=& \fft{9n\, \omega}{2\pi}\,
\int \fft{\sinh^2\rho\, d\rho}{\sqrt{\kappa^2\, \cosh^2\rho - \omega^2\,
\sinh^2\rho - c^2/\sinh^2\rho}}\nn\\
\eea
%%%%%
Thus $E$ and $J$ can
also be expressed as hypergeometric functions, given by 
%%%
\bea E&=&
\fft{9n\, \kappa\, \pi\, \sqrt{y_+}}{2\sqrt{\omega^2-\kappa^2}} \,
\FF2[\ft12, -\ft12, 1; \lambda]\,,\nn\\ \fft{E}{\kappa} - \fft{S}{\omega}
&=& \FF2[\ft12, \ft12, 1; \lambda]\,.  
\eea 
%%%%

     Not surprisingly, the result is the same as that for a string
rotating in the AdS$_5\times S^5$ background.  This is because we have
considered a solution in which $\theta_1=0$, which is an AdS$_5$ slice
of AdS$_6$.  From the AdS/CFT point of view, the ${\cal N}=1$ $D=4$
Yang-Mills can be obtained from a Kaluza-Klein truncation of the
${\cal N}=2$ conformal field theory in $D=5$.  In order to probe the
non-trivial properties of the $D=5$ conformal field theory, it would
be necessary to look for solutions with $\theta_1\ne0$.  However, what
we have illustrated implies that there are common features in the
leading Regge trajectories for the two theories, whilst in their large
R-charge expansions the two theories diverge more significantly.

\section{Conclusions}

      In this paper, we have contructed spinning and rotating string
solutions in the massive type IIA warped AdS$_6\times S^4$ background.
String theory on such a background is expected to be dual to an 
${\cal N}=2$, $D=5$ super-conformal Yang-Mills theory.

     In the case of a string spinning purely in the AdS$_6$, the
gravitional repulsion from the equatorial boundary implies that the
string can only lie at the north pole of the $S^4$.  Besides this, the
warp factor plays no essential role.  The resulting $E$ and $S$
relation is the same as that obtained in the AdS$_5\times S^5$
background, implying the same behaviour of the leading Regge
trajectories of the two theories.

        In the case of a string rotating in the $S^4$, the warp factor
plays a more significant role.  First, it rules out a string rotating
only in the $S^3$ that foliates the $S^4$, which implies that the
analogous energy vs. R-charge relation in the $D=4$ Yang-Mills theory
associated with AdS$_5\times S^5$ is absent in the $D=5$ Yang-Mills
theory.  We obtained analytic solutions for the rotating string
extending in the latitude coordinate of the $S^4$.  Owing to the
presence of the warp factor, the energy and angular momentum relation
is qualitatively different for the long string at the sub-leading
order, in comparison to the relation in AdS$_5\times S^5$..

    We also studied Penrose limits of the AdS$_6\times S^4$
background.  In one limit, we obtained a massive string theory with
masses that depend upon the worldsheet time coordinate $x^+$ in the
light-cone gauge.  The range of the time coordinate is restricted,
owing to the presence of the warp factor, which diverges at a finite
value of $x^+$.  At the same time, the string coupling constant
becomes infinite, which may signal a breakdown of the validity of the
solution.  A second, inequivalent Penrose limit was also constructed,
which yields a time-independent interacting string theory.  The two
Penrose limits arise from schemes that would have given identical
results in a situation such as AdS$_5\times S^5$ where there is no
warp factor.

    In comparison to the AdS$_5\times S^5$ solution of the type IIB
theory, the AdS$_6\times S^4$ solution of the massive type IIA theory
exhibits a number of undesirable features, all of which stem from the
warp factor which becomes singular on the equator of the $S^4$.  In
particular, the string probe senses this singularity, in any motion
that approaches the equator.  Interestingly, however, a 4-brane probe
is insensitive to the warp factor.  In other words, if one makes a
Weyl rescaling of the metric to the ``4-brane frame,'' then it becomes
purely a direct product of AdS$_6$ and $S^4$.  By definition, the rescaling
of the metric to the 4-brane frame is such that the dual field strength
$F_\6 \equiv e^{\ft12\phi}\, {*F_\4}$, to which a fundamental 4-brane 
couples, arises in the massive type IIA supergravity Lagrangian with
the same dilaton coupling as for the $\sqrt{-g}\, R$ term, namely
%%%
\be
{\cal L} = e^{-\ft25\phi}\, \sqrt{-g}\, (R - \ft{1}{6!} F_\6^2 + 
\cdots)\,.
\ee
%%%
Thus the 4-brane metric is related to the Einstein metric by $ds_{\rm
4-brane}^2 = e^{\ft1{10}\phi}\, ds_{\rm Ein}^2$.  We then see from
(\ref{ads6s4}) that in the 4-brane frame, the metric becomes a
direct product of AdS$_6 \times S^4$.  It would seem that 4-branes
might therefore be more natural candidates for probing the geometry of
the background.  Unfortunately, however, the necessity of wrapping a
spinning or rotating 4-brane in all the available $U(1)$ circle
isometries of AdS$_6$ and $S^4$ appears to exclude the possibility of
obtaining simple solutions.

         The warped AdS$_6\times S^4$ background of the massive type
IIA supergravity, as a near-horizon geometry of the D4-D8 system, may
provide the simplest arena for studying super-conformal field theories
beyond $D=4$.  The preliminary analysis of our paper suggests that
there are many common features with the $D=4$ Yang-Mills theory.  The
new features, other than dimensionality, are principally due to the
warped nature of the AdS$_6\times S^4$ product.  In particular, this
leads to energy vs. R-charge relationships that differ from those in
the $D=4$ Yang-Mills theory.  It would be of interest to study this
further from the standpoint of the five-dimensional superconformal
field theory.

\section*{Acknowledgments}

   We are grateful to Carlos Nunez and Per Sundell for useful discussions.

\end{document}